\documentclass[preprint,review,3p,times]{elsarticle}

\usepackage[colorlinks,urlcolor=blue,linkcolor=blue,citecolor=blue]{hyperref}

\usepackage{color,array}
\usepackage{adjustbox}

\usepackage{pdflscape} 
\usepackage{graphicx}
\usepackage{float}
\usepackage[linewidth=1pt]{mdframed}
\usepackage{afterpage}
\usepackage{longtable}
\usepackage{url}
\usepackage{caption}
\usepackage{epstopdf}
\usepackage{amsmath}
\usepackage{amssymb}
\usepackage{xcolor}
\usepackage{rotating}
\usepackage{booktabs}
\usepackage[ruled,vlined]{algorithm2e}

\usepackage{pgfplots}
\pgfplotsset{compat=1.18}
\usepackage{tikz}

\usepackage[colorinlistoftodos]{todonotes}
\presetkeys{todonotes}{color=blue!20}{}

\usepackage{multirow}

\usepackage{array}
\newcolumntype{C}{ >{\centering\arraybackslash} m{1.0cm} }
\newcolumntype{Q}{ >{\centering\arraybackslash} m{3 cmcm} }
\newcolumntype{M}[1]{>{\centering\arraybackslash}m{#1}}
\newcolumntype{P}[1]{>{\centering\arraybackslash}p{#1}}
\usepackage{epstopdf}

\begin{document}
	\pagenumbering{arabic}
		\begin{frontmatter}
	\title{LAMP-PRo: Label-aware Attention for Multi-label Prediction of DNA- and RNA-binding Proteins using Protein Language Models}
	\author[1, *]{Nimisha Ghosh\corref{mycorrespondingauthor}}%
	\ead{nimishaghosh@snuchennai.edu.in}
	\author[1]{Dheeran Sankaran}
	\ead{dheeran24110056@snuchennai.edu.in}
	\author[1]{Rahul Balakrishnan Adhi}
	\ead{rahul24110299@snuchennai.edu.in}
	\author[1]{Sharath S}
	\ead{sharath24110418@snuchennai.edu.in}
	\author[1]{Amrut Anand}
	\ead{amrut24110298@snuchennai.edu.in}

		\address[1]{Department of Computer Science and Engineering, Shiv Nadar University Chennai, Tamil Nadu, India}
	
	\begin{abstract}
		Identifying DNA- (DBPs) and RNA-binding proteins (RBPs) is crucial for the understanding of cell function, molecular interactions as well as regulatory functions. Owing to their high similarity, most of the existing approaches face challenges in differentiating between DBPs and RBPs leading to high cross-prediction errors. Moreover, identifying proteins which bind to both DNA and RNA (DRBPs) is also quite a challenging task. In this regard, we propose a novel framework viz. LAMP-PRo which is based on pre-trained protein language model (PLM), attention mechanisms and multi-label learning to mitigate these issues. First, pre-trained PLM such ESM-2 is used for embedding the protein sequences followed by convolutional neural network (CNN). Subsequently multi-head self-attention mechanism is applied for the contextual information while label-aware attention is used to compute class-specific representations by attending to the sequence in a way that is tailored to each label (DBP, RBP and non-NABP) in a multi-label setup. We have also included a novel cross-label attention mechanism to explicitly capture dependencies between DNA- and RNA-binding proteins, enabling more accurate prediction of DRBP. Finally, a linear layer followed by a sigmoid function are used for the final prediction. Extensive experiments are carried out to compare LAMP-PRo with the existing methods wherein the proposed model shows consistent competent performance. Furthermore, we also provide visualization to showcase model interpretability, highlighting which parts of the sequence are most relevant for a predicted label. 
		The original datasets are available at http://bliulab.net/iDRBP\_MMC and the codes are available at https://github.com/NimishaGhosh/LAMP-PRo.
	\end{abstract}

\end{frontmatter}

	\section{Introduction}
	To understand complex cellular operations, it is imperative to have knowledge of the functional roles being executed by nucleic acid binding proteins (NABP); broadly classified as DNA-binding proteins (DBPs) and RNA-binding proteins (RBPs). The key functionalities of DBPs and RBPs include transcription and translation regulation, DNA replication and repair, chromatin organisation, RNA splicing etc. Thus, analysis of these proteins may unveil the fundamental mechanisms of several diseases like cancer, neurological disorders, myopathies etc. Thus, classifying such proteins based on their functional characteristics enable the researchers to understand the molecular processes of a cell. However, there are several challenges that are faced while executing such classification. First of all, DBPs and RBPs have very similar structure and highly correlated in terms of function and evolution as well, making it highly difficult to distinguish them, thus leading to cross-prediction problem; DBP is identified as RBP and vice-versa. Second, most of the works do not explore the possibility of designing a model for multi-label learning to identify the relationship between labels. In this regard, there are some works like~\cite{Du2023, Prabhu2025} who have applied deep learning architecture as well as multi-label learning for such prediction.
	
	To identify DBPs and RBPs, high throughput screening techniques can be considered to be an important step~\cite{Cui2022}. Many other methods have also been employed such as nuclear magnetic resonance spectroscopy~\cite{Douglas2007}, X-ray crystallography~\cite{Jaiswal2015}, etc. These methods can provide a comprehensive understanding of how proteins bind to DNA and RNA. However, they are time-consuming and expensive, leading to the development of computational methods. In this regard, several works have come up in the past few years but mostly they consider DBP and RBP recognition tasks separately; very few of them address both. For DBP prediction, in~\cite{Kumar2007}, the authors have used evolutionary information along with Support Vector Machine (SVM) wherein they achieved an accuracy of 86.62\%. Deep-WET proposed in~\cite{Mahmud2024} uses feature encoding schemes like Global Vectors, Word2Vec and fastText for protein sequence, the results of which are combined and passed through Differential Evolution (DE) algorithm. Finally, the optimal feature subset learned from SHapley Additive exPlanations (SHAP) are provided as input to Convolutional Neural Network (CNN) for the final prediction. This led to an accuracy of 78.08\%, MCC of 0.550 and AUC of 0.805. Ahmad et al.~\cite{Ahmed2024} have proposed StackDPP which utilises Random Forest (RF)  and Recursive Feature Elimination with Cross-validation (RFECV) to select best features and eventually uses a stacking archutecture for the final prediction of DBPs. Independent testing on the PDB186 dataset fetched an accuracy of 93\%. LBi-DBP proposed by Zeng et al.~\cite{ZENG2024} uses five sequence feature sources like position specific scoring matrix (PSSM), Hidden Markov Model Profile (HMM), Predicted Secondary Structure Probability Matrix (PSSPM), Predicted Solvent Accessibility Probability Matrix (PSAPM) and Predicted Probabilities of DNA-Binding Sites (PPDBS) and subsequently applies BiLSTM to extract the protein sequence context information. This information is then combined with five pseudo sequence order (SO) features and provided as input to an MLP-based neural network module to predict DBP. On UniSwiss-Tst and PDB2272 datasets, LBi-DBP achieves MCC values of 0.762 and 0.574.

	In~\cite{Kumar2011}, Kumar et al. have considered evolutionary information in the form of  PSSM for the encoding. Further, they have used SVM to distinguish between RBP and non-binding proteins with an MCC of 0.66. Ma et al.~\cite{Ma2015} have proposed PRBP which uses RF along with an RNA-binding residue to predict RBPs. PRBP achieves an accuracy of 85.61\%, MCC of 0.67 and AUC of 0.90. Zhang et al.~\cite{Zhang2016} have also explored the detection of RBPs by combining evolutionary information and physicochemical properties of protein sequences along with SVM. In this regard, they have achieved an AUC of 0.975 and MCC of 0.814. This work has been further improved in~\cite{Zheng2018} by introducing CNN where they achieved an MCC of 0.82. More recently, Yan et al. have proposed Seq-RBPPred~\cite{Yan2024} which uses feature representation from biophysical properties and hidden-state features derived from protein sequences by employing Protr, UniRep, SeqVec and ESM-1b. Henceforth, they have used XGBoost on these features to predict RBPs. This  resulted in an accuracy of 92\% and MCC of 0.75. 
	
	Some notable works which have addressed both DBPs and RBPs are described next. Zhang et al.~\cite{Zhang2021} have proposed DeepDRBP-2L which uses CNN and LSTM for the prediction of DBP and RBP. Although the prediction accuracy of DBP is quite good, RBP is relatively low. Overall, the model achieves a total accuracy of 80\%.  Wu et al.~\cite{Wu2024} employ two layers of CNN and one layer of LSTM to predict DBPs and RBPs resulting in the total prediction accuracy of 81\%. iDRBP\_MMC proposed in~\cite{ZHANG2020}  uses multi-label learning strategy during training process to reduce cross-prediction between DNA-binding proteins and RNA-binding proteins. It incorporates motif-based CNN and traditional CNN to capture special patterns of nucleic acid binding proteins. In another work, Wang et al.~\cite{Wang2023} have proposed a hierarchical ensemble learning method to predict DNA-binding and RNA-binding proteins. It fuses 60 base models which integrates multiple tasks, feature sets and classification algorithms for such prediction. However, a major limitation lies in their scalability, as the process of designing, training, and integrating multiple models reduces both interpretability and transparency.
	DMJL~\cite{Du2023} uses a deep multi-label joint learning framework that exploits the relationships between multiple labels and binding proteins. First, a multi-label variant network is constructed to capture hidden information across multiple scales. Next, a multi-label Long Short-Term Memory (multiLSTM) module is employed to uncover potential dependencies among the labels. Finally, the calibrated hidden features from the variant network are integrated into different levels of joint learning, enabling the multiLSTM to more effectively model the correlations between the labels. However, DMJL relies on a shared feature representation and does not provide label-specific attention nor explicit modeling of biologically meaningful label interactions. This may potentially lead to suboptimal discrimination of complex label combinations and occasional logical inconsistencies. SERCNN proposed by Prabhu et al.~\cite{Prabhu2025}  is a multilabel classification model, which has deep neural networks with residual connections and squeeze-and-excitation attention mechanisms for the prediction of DBPs, RBPs as well as DRBPs. Although, the results are quite competent, this work suffers from some drawbacks.
	SERCNN treats DBP and RBP prediction as independent binary tasks, thus lacking the capability to explicitly model co-binding. These drawbacks have motivated us to propose a attention based model viz. LAMP-PRo which combines label-aware  and cross-label attention thus enabling both label-specific feature learning as well as dynamic modelling of dependencies between DBP and RBP for improved DRBP recognition. To summarise, the main contributions of this work are as follows:
	\begin{enumerate}
		\item A novel framework combining label-aware and cross-label attentions, enabling LAMP-Pro to capture both label-specific evidence (DBP, RBP and Non-NABP) and label dependencies (DBP-RBP for DRBP).
		\item Introduction of gated residual connections between CNN and MHSA as well as between label-aware and cross-label attention to adaptively regulate local–global feature fusion and stabilize training.
		\item Performing extensive ablation studies to highlight the importance of different types of attention components in LAMP-PRo.
		\item Making the code publicly available so that the research community can benefit from it (Neither SERCNN nor DMJL have provided any public repository for their codes (Although there is a link provided for DMJL, it is currently not functional)).
	\end{enumerate}
	\section{Materials and Methods}
	In this section, we discuss the datasets used in this work followed by the pipeline of the work.
	\subsection{Datasets}
	In this work, we have used the same datasets as used by Zhang et al.~\cite{ZHANG2020}. The statistics of the training and test dataset is given in Table~\ref{tab1}.
	\begin{table}[H]\scriptsize
		\centering
		\begin{tabular}{llllll}
			\hline
			Dataset & DBPs  & RBPs & DRBPs & Non-NABPs & Total \\
			\hline
			Training Dataset & 3846 & 2616 & 329 &  4175 & 10966\\
			EZL Dataset & 2226 & 1777 & 0 & 0 & 4003\\
			TEST474 & 175 & 68 & 8 & 223 & 474\\
			PDB255 & 93 & 70 & 0 & 92 & 255\\
			DRBP206 & 0 & 0 & 103 & 103 & 206\\\hline
			
		\end{tabular}
		\caption{Summary of Training and Test Datasets}
		\label{tab1}
	\end{table}
	As reported in the table, the total number of proteins that bind to DNA (DBPs), RNA (RBPs), both DNA and RNA (DRBPs) and that do not bind to any nucleic acid (Non-NABPs) are 3846, 2616, 329 and 4175 respectively leading to a total of 10966 proteins in the training dataset. In this work, DBPs, RBPs and Non-NABPs are treated as protein labels, leading to a multi-label learning scenario that focuses on three fixed class labels; [1, 0, 0] for DBP, [0, 1, 0] for RBP and [0, 0, 1] for non-NABP. By default, DRBP which is a combination of both DBP and RBP is thus considered as [1, 1, 0] and not taken as a different class label. Thus, the class labels of both training and test datasets are consistent.
	
	\subsection{Pipeline of the Work}
\begin{figure*}
		\centering
		\includegraphics[height=3in,width=6.0in]{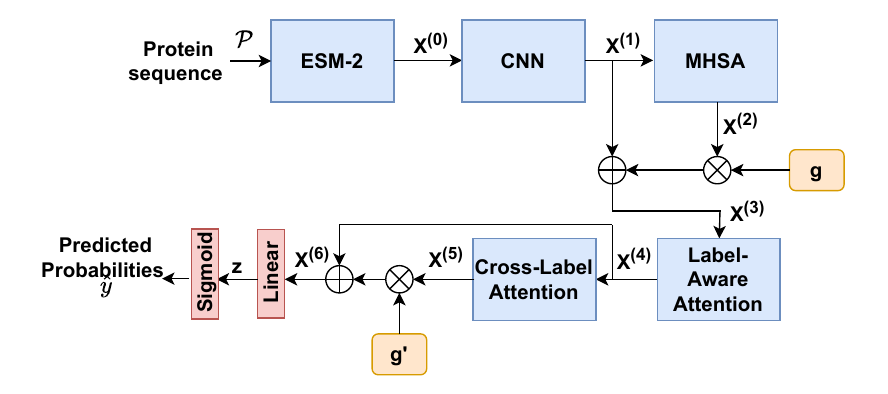}
		\caption{Pipeline of LAMP-PRo. Initially, ESM-2 is used to extract the embeddings from the input protein sequences. Next, CNN is applied to extract local features while MHSA captures the global features. Later label-aware attention learns different attentions per label for DBP, RBP and non-NABP. Cross-label attention further models dependencies between DBP and RBP. Finally, a linear layer followed by a sigmoid function produces multi-label probabilities.}
		\label{fig:pipeline}
	\end{figure*}

	The pipeline of the proposed work LAMP-PRo is given in Fig.~\ref{fig:pipeline}. The model has five distinct components which are described next in details.
	\subsubsection{Protein Embedding}
	The protein sequences are first provided as input to ESM-2 pre-trained protein language model, thereby leveraging the knowledge learned from massive pretraining on protein sequences. This transforms the protein sequences into vector representations which can be subsequently used as input to the next components of the model. ESM-2 has different model sizes ranging from 8M to 15B parameters. However, due to computational constraints, we have chosen the ESM-2 model with 150M parameters. Thus, given a protein sequence of length $L$, the embedded matrix has a size of $L\times 640$. Let us consider an input protein sequence $\mathcal{P} = \{p_1,p_2,\dots,p_L\}$, where $L$ is the sequence length. For a given input sequence $\mathcal{P}$, ESM-2 generates emebeddings $X^{(0)}$ given by:
	\begin{eqnarray}
		X^{(0)} = ESM-2 (\mathcal{P}) \in \mathbb{R}^{L \times d_e} 
	\end{eqnarray}
	Here, $d_e$ = 640, the embedding dimension.
	\subsubsection{CNN}
	The embeddings from ESM-2 ($X^{(0)}$) are passed through a 1D CNN component for high-order local pattern extraction. This component detects conserved patterns important for binding patterns. Moreover, it also helps compressing the long sequence from ESM-2, making the model more computationally efficient. It also help reduce the sequence length (via stride) to speed up the subsequent attention layers. In this regard, a single convolutional operation (Conv1D) is used followed by batch normalization (BN), gaussian error linear unit (GELU) and dropout operation. These steps can be written as:
	\begin{equation}
		\begin{aligned}
			X^{(1)} &= Dropout(GELU(BN(Conv1D(X^{(0)})))) \\
			& \in \mathbb{R}^{L' \times d_c} 
		\end{aligned}
	\end{equation}
	where, L' is the new sequence length after convolution and $d_c$ is the CNN output channels
	\subsubsection{Multi-head Self-Attention}
	The output from the 1D CNN block ($X^{(1)}$) is then provided as an input to the multi-head self-attention block. This component ensures that each position in the sequence attend to others, thereby capturing global dependencies. Self-attention provides the flexibility to focus on important residues irrespective of the distance while multi-head mechanism allows learning of diverse relation patterns in the sequence. For each self-attention head $i = 1,\dots,h$,
	\begin{eqnarray}
		head_i = Attention (Q_i,K_i,V_i) = softmax (\frac{Q_iK_i^T}{\sqrt{d_i}})V_i
	\end{eqnarray}
	where $Q_i$ = $X^{(1)}W_i^Q$,  $K_i$ = $X^{(1)}W_i^K$,  $V_i$ = $X^{(1)}W_i^V$.
	Finally, the multi-head self-attention can be given as:
	\begin{equation}
		\begin{aligned}
			X^{(2)} &= \text{MHSA}(X^{(1)}) \\
			&= \text{Concat}(\text{head}_1, \dots, \text{head}_h) W^O 
			\in \mathbb{R}^{L' \times d_c}
		\end{aligned}
	\end{equation}
	
	Additionally, to capture both local motif-level patterns and long-range dependencies, the CNN and the self-attention features are combined with a gated residual ( $g = \phi(\mathcal{G}) \in (0,1)$, where $\phi$ denotes a sigmoid activation function for gating). The gate allows the network to adaptively learn and control how much of the global context should be included while the residual connection preserves the original local features for stable training. Mathematically, it can be expressed as:
	\begin{eqnarray}
		X^{(3)} = X^{(1)} \oplus g \odot X^{(2)} \in \mathbb{R}^{L' \times d_c}
	\end{eqnarray}
	Here, $\oplus$ and $\odot$ denote element-wise addition and multiplication respectively. 
	\subsubsection{Label-aware Attention}
	For labels such as DBP, RBP and Non-NABP, this component computes a dedicated attention-weighted summary for each label over the features refined by the multi-head self-attention mechanism. Unlike standard pooling (max/average), label-aware attention learns different attention patterns per label, thus capturing label-specific evidence from the sequence and enabling multi-label awareness. 
	
	Let the number of labels be $C$ = 3 (DBP, RBP and Non-NABP). Let the learned label embeddings be denoted by $E_{label} \in \mathbb{R} ^{C \times d_c}$, while query, key and value are denoted by $\mathcal{Q} = E_{label}$, $\mathcal{K} = X^{(3)}$ and $\mathcal{V} = X^{(3)}$ respectively. The attention output is finally calculated as:
	\begin{eqnarray}
		X^{(4)} = softmax (\frac{\mathcal{Q}\mathcal{K}^T}{\sqrt{d}})\mathcal{V}\in \mathbb{R}^{C \times d_c}
	\end{eqnarray}
	
	\subsubsection{Cross-Label Attention}
	The previous component focuses on labels like DBP, RBP and non-NABP. However, this may lead to misclassification in sequences for DRBP labels which considers the joint presence of both DBP and RBP. To explicitly capture dependencies among DBP and RBP where DRBP label is present and to improve its embeddings, a cross-label attention module operating on the label-aware embeddings is introduced. Unlike conventional self-attention across tokens, queries, keys and values are the label representations. Furthermore, the attention mechanism is restricted by a label mask $\mathcal{M} \in \{0,1\}^{{C \times C}}$ which encodes which label-to-label interactions are permitted. For instance, DBP and RBP labels can attend to each other in the DRBP case. Importantly, DRBP is not predicted as a separate logit but inferred at the output stage from the co-activation of DBP and RBP probabilities. 
	
	The label mask can be depicted as:
	\begin{equation}
		\begin{aligned}
			M_{i,j} =
			\begin{cases}
				0, & \text{if label $i$ can attend to $j$} \\
				-\infty, & \text{otherwise}
			\end{cases}
		\end{aligned}
	\end{equation}
	The masked attention for each cross-attention head $t$ is given by:
	\begin{eqnarray}
		X^{(5)} = softmax (\frac{\mathcal{Q}_{t}\mathcal{K}_{t}^T}{\sqrt{d}} + M)\mathcal{V}_{t}\in \mathbb{R}^{C \times d_c}
	\end{eqnarray}
	where $\mathcal{Q}_{t}$ = $X^{(4)}W_{t}^\mathcal{Q}$,  $\mathcal{K}_{t}$ = $X^{(4)}W_{t}^\mathcal{K}$ and $\mathcal{V}_{t}$ = $X^{(4)}W_{t}^\mathcal{V}$. 
	
	Furthermore, a gated residual connection ($ g' = \phi(\mathcal{G'}) \in (0,1)$, where $\phi$ denotes a sigmoid activation function for gating) is introduced to adaptively control how much cross-label information should be used. This design stabilises training and improves multi-label representation learning, particularly for DRBP sequences where dependencies between DBP and RBP must be selectively integrated. This step is particularly important where in the presence of DRBP sequences, the gate can learn to allow stronger cross-label interactions, improving DRBP detection. 
	The final output is given by:
	\begin{eqnarray}
		X^{(6)} = X^{(4)} \oplus g' \odot X^{(5)} \in \mathbb{R}^{C \times d_c} 
	\end{eqnarray}

	\subsubsection{Prediction Layer}
	The prediction layer comprises of a linear layer and a sigmoid function  for the final output. The linear layer projects the attended features to a logit space for prediction while sigmoid function is responsible for multi-label probability output. It converts each of the 3 logits for DBP, RBP and Non-NABP into probabilities $\in [0,1]$.  It is important to note that DRBP is concluded from the co-activation of DBP and RBP outputs. These steps can be outlined as:
	\begin{eqnarray}
		\hat{y} = \sigma(z) \in [0,1]^C 
	\end{eqnarray}
	where, $z = Linear(X^{(6)}) \in \mathbb{R}^C$ is the logit and $\hat{y}$ is the final prediction probability for DBP, RBP and Non-NABP while DRBP is not directly predicted but inferred from DBP and RBP.
	
	\subsection{Model Training}
	For training purpose, Binary Cross-Entropy  loss is applied:
	\begin{eqnarray}
		\mathcal{L}_{main} = BCE(\sigma(z),y)
	\end{eqnarray}
	where, $y$ is the actual label. Additionally, invalid label penalty is applied on predicted probabilities $\hat{y}$ as:
	\begin{eqnarray}
		\mathcal{L}_{penalty} = InvalidLabelPenalty(\hat{y})
	\end{eqnarray}
	The total loss is thus:
	\begin{eqnarray}
		\mathcal{L}_{total} = \mathcal{L}_{main} + \lambda \cdot \mathcal{L}_{penalty}
	\end{eqnarray}
	where, $\lambda$ is a hyperparameter controlling the trade-off between the main classification loss and the penalty. 
	Since this work is based on multi-label classification, some label combinations may be logically inconsistent. For example, DBP and Non-NABP cannot occur together. Thus, invalid label penalty is incorporated as part of the loss calculation to explicitly ensure that impossible label combinations are discouraged during training.
	
	The pseudo-code for LAMP-PRo is given in Algorithm~\ref{algo} 
	
	\begin{algorithm}
		\caption{Pseudo-Code of LAMP-PRo}
		\label{algo}
		\KwIn{Protein sequence $\mathcal{P}$, labels $y \in \{0,1\}^3$ (DBP, RBP, Non-NABP)}
		\KwOut{Predicted probabilities $\hat{y} \in (0,1)^3$}
		\BlankLine
		
		\textbf{Step 1: Embedding extraction} \\
		\quad Obtain sequence embeddings $X^{(0)} = \text{ESM-2}(\mathcal{P})$
		
		\textbf{Step 2: Local–global feature extraction} \\
		\quad $X^{(1)} = \text{CNN}(X^{(0)})$ \tcp{local motif features} 
		\quad $X^{(2)} = \text{MHSA}(X^{(1)})$ \tcp{global dependencies} 
		\quad $X^{(3)} = X^{(1)} \oplus g \odot X^{(2)}$ \\
		\quad where $g = \sigma(\mathcal{G})$ (gated residual)
		
		\textbf{Step 3: Label-aware attention} \\
		\quad Compute label-specific embeddings $X^{(4)} = \text{LAA}(X^{(3)})$ for $\ell \in \{\text{DBP, RBP, Non-NABP}\}$
		
		\textbf{Step 4: Cross-label attention} \\
		\quad Apply restricted attention among labels using mask $M$ \\
		\quad $X^{(5)} = \text{CLA}(X^{(4)}, M)$\\
		\quad $X^{(6)} =  X^{(4)} \oplus g' \odot X^{(5)}$ \\
		\quad where $g' = \phi(\mathcal{G'})$ (gated residual)\\
		\textbf{Step 5: Prediction} \\
		\quad $\hat{y} = \sigma(\text{Linear}(X^{(6)}))$
		
		
		
	\end{algorithm}
	
	\subsection{Performance Evaluation}
	The first and foremost performance evaluation metric that is considered in this work is area under the receiver operating characteristic (ROC)
	curve (AUC). The value of AUC ranges between 0 and 1 where a higher value indicates a better classification ability. As both DBP and RBP are part of all the datasets used in this work, there is a very high chance of cross-prediction. In this regard, AURC curve can be considered to be another very important metric. Specifically, 1-AURC demonstrates the classifier's ability to manage cross-prediction issues. In this case, a larger value means that the model has a very strong ability to mitigate cross-prediction.
	
	\section{Results and Discussion}
	The results obtained from LAMP-PRo are elaborately discussed in this section. 
	\subsection{Hyperparameter Settings}
	The different hyperparameters considered during training are listed in Table~\ref{tab:hyper}. The final values for the different parameters are chosen after extensive experiments. We have used early stopping regularization technique to prevent overfitting. The maximum number of epochs we have considered in this case is 15. If for two consecutive epochs, the model does not show improved performance on AUC-ROC, the training stops.
	
	\begin{table}[H]\tiny
		\centering
		\begin{tabular}{lcc}\\\hline
			Hyperparameters   & Choices & Chosen Value  \\\hline
			Learning Rate& 1e-2, 1e-3, 1e-4 & 1e-4 \\
			Number of filters (CNN) & 64, 128, 256, 512 & 256\\
			Number of Heads (MHSA) & 2, 4, 6 & 4\\
			Number of Heads (CLA) & 2, 4, 6  & 2\\
			Loss & Binary cross-entropy, Focal loss & Binary cross-entropy\\
			Batch size & 32, 64 & 32\\
			$\lambda$ (hyperparameter for invalid label penalty) & 0.1, 0.2, 0.3 & 0.1\\\hline
		\end{tabular}
		\caption{Hyperparameter Settings}
		\label{tab:hyper}
	\end{table}
	
	In order to determine which ESM-2 model is suitable for embedding, we have performed extensive experiments on ESM-2-8M, ESM-2-35M and ESM-2-150M. Due to our computational resource constraint, we could not perform on larger models. As expected ESM-2-150M model shows the most competent performance among all and is chosen as the final embedding model. All the experiments are performed on 2 A40 GPUs.
	
	\subsection{Ablation Studies}
	LAMP-PRo has several components like ESM-2, CNN, Multi-Head self attention, Label-aware attention and Cross-Label attention which have distinct functionalities that contribute to the competent performance of the model. In order to showcase their individual contributions, ablation studies are performed on the four independent test datasets with different variants.
	\begin{itemize}
		\item ESM-2 + CNN + LAA + CLA + Linear + Sigmoid (Variant 1): In this variant, the multi-head self attention component is removed. 
		\item ESM-2 + CNN + MHSA  + CLA + Linear + Sigmoid (Variant 2): In this variant, label-aware attention is removed to show its contribution towards multi-label classification.
		\item ESM-2 + CNN + MHSA  + LAA + Linear + Sigmoid (Variant 3): Cross-label attention is removed in this variant in order to showcase its importance in predicting DRBPs.
		\item ESM-2 + CNN + MHSA  + LAA + CLA + Linear + Sigmoid (Variant 4): In this variant, the full model is retained but without \textit{g}.
		\item ESM-2 + CNN + MHSA  + LAA + CLA + Linear + Sigmoid (Variant 5): In this variant, the full model is retained but without \textit{g'}.
		\item LAMP-PRo (Variant 6): This variant depicts our full model.
	\end{itemize}
	\begin{table*}[ht]\scriptsize
		\centering
		\caption{Ablation Study of the four variants on the independent test datasets TEST474, PDB255, EZL and DRBP206 datasets for DNA and RNA binding prediction}
		\begin{tabular}{l|cc|cc|cc|cc|cc|cc|c}
			\hline
			\multirow{2}{*}{\textbf{Method}} & \multicolumn{4}{c|}{\textbf{TEST474}} & \multicolumn{4}{c|}{\textbf{PDB255}} & \multicolumn{4}{c|}{\textbf{EZL Dataset}} & \textbf{DRBP206}\\\cline{2-14}
			& \multicolumn{2}{c|}{DNA Binding} & \multicolumn{2}{c|}{RNA Binding} & \multicolumn{2}{c|}{DNA Binding} & \multicolumn{2}{c|}{RNA Binding} & \multicolumn{2}{c|}{DNA Binding} & \multicolumn{2}{c|}{RNA Binding} & DRBP\\\cline{2-14}
			& AUC & 1-AURC & AUC & 1-AURC & AUC & 1-AURC & AUC & 1-AURC & AUC & 1-AURC & AUC & 1-AURC & AUC\\\hline
			Variant 1 & 0.96 & 0.94 & 0.87 & 0.91 & 0.80 & 0.79 & 0.75 & 0.73 & 0.93 & 0.93 & 0.95 & 0.95 & 0.94\\
			Variant 2 & 0.50 & 0.50 & 0.50 & 0.50 & 0.50 & 0.50 & 0.50 & 0.50 & 0.50 & 0.50 & 0.50 & 0.50 & 0.50\\
			Variant 3 & 0.97 & 0.94 & 0.90 & 0.92 & 0.79 & 0.76 & 0.75 & 0.76 & 0.94 & 0.94 & 0.95 & 0.95 & 0.70 \\
			Variant 4 & 0.98 & 0.96 & 0.90 & 0.93 & 0.82 & 0.80 & 0.79 & 0.80 & 0.95 & 0.95 & 0.96 & 0.96& 0.97\\
			Variant 5& 0.98 & 0.95 & 0.86 & 0.90 & 0.81 & 0.78 & 0.78 & 0.78 & 0.95 & 0.95 & 0.96 & 0.95& 0.99\\
			Variant 6 (LAMP-PRo)& 0.98 & 0.97 & 0.90 & 0.95 & 0.80 & 0.78 & 0.81 & 0.83 & 0.96 & 0.96 & 0.96 & 0.96 & 0.96 \\
			
			\hline
		\end{tabular}
		\label{tab:ablation}
	\end{table*}
	
	\begin{figure*}
		\centerline{
			\includegraphics[height=2.5in,width=3.0in]{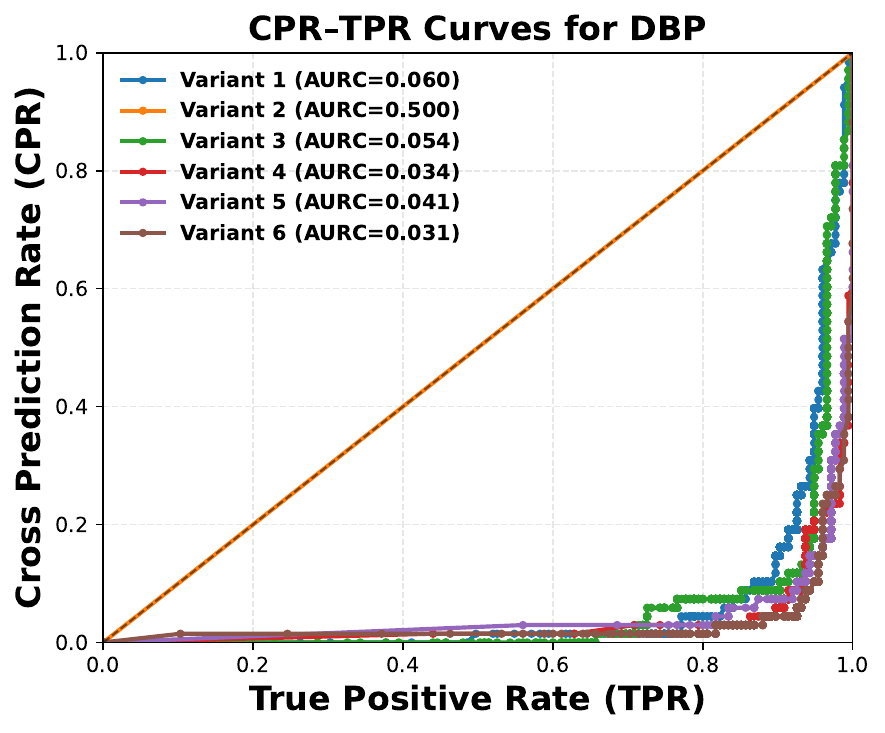}
			\includegraphics[height=2.5in,width=3.0in]{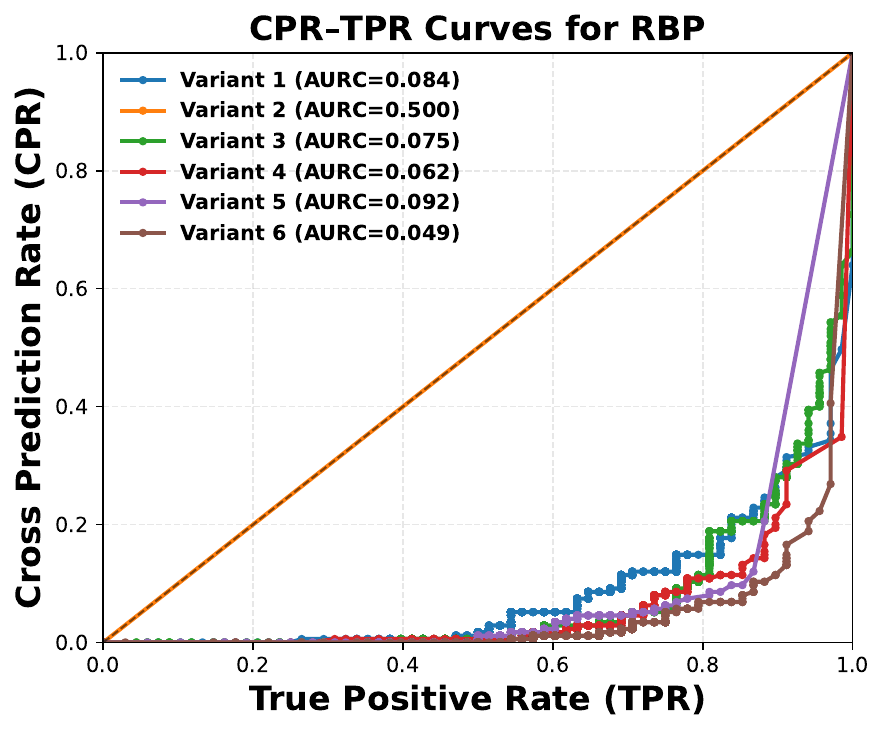}
		}
		\centerline{(a)\hspace{80mm}(b)}
		\caption{Cross-Prediction performance of the different variants on TEST474 dataset}
		\label{Ablation_figure}
	\end{figure*}
	
	As can be observed from the results of Table~\ref{tab:ablation}, the removal of multi-head self attention in Variant 1 shows somewhat reduction in performance for most of the datasets. This is because MHSA captures global dependencies and contributes to improved performance.
	It is interesting to note that all the variants except Variant 2 show quite strong performance for all the datasets. This can be attributed to the fact that Variant 2 does not have label-aware attention which contributes to the multi-label awareness, thereby showing very poor performance. On the other hand, for DRBP206 dataset, Variant 3 also shows poor performance. This can be due to the fact that cross-label attention is removed in Variant 3 and there are only DRBPs in DRBP206 dataset. Thus cross-label attention should have the maximum contribution; due to its absence the variant shows poor performance. 
	The effects of removal of gating mechanisms are shown through Variants 4 and 5. Such mechanisms exhibit dataset-specific effects. On datasets such as TEST474, PDB255 and EZL gates stabilize performance, especially for RNA predictions. However, for DRBP206 dataset, gates suppress informative overlap, so removing them boosts AUC. MHSA-CNN fusion gate ($g$) is designed to balance local and global representations. On diverse datasets such as TEST474, this balance improves generalization by preventing over-reliance on one feature type. However, on the DRBP206 dataset, where the task is simpler and strongly local features are sufficient, the additional gating acts as unnecessary regularization. This suppresses useful signals and slightly reduces performance, explaining why removal of $g$ improves DRBP results. On the other hand, the gate ($g'$) between label-aware and cross-label attention is  designed to stabilize training by controlling cross-label feature flow. On diverse datasets such as TEST474, where DBP, RBP and DRBP co-exist, the gate prevents spurious cross-talk and improves generalization. However, in DRBP206, where only DRBP v/s Neither labels are present, the cross-label interactions are inherently simple. In this setting, the gate effectively acts as an over-regularizer which is initialized to suppress interactions and thus it may downweight useful DRBP features, thereby slightly reducing the performance. Removing the gate therefore improves DRBP detection by allowing full integration of label-aware and cross-label representations. 
	The inclusion of gates also improves cross-prediction performance which is further evident from Figure~\ref{Ablation_figure}. Moreover, as shown in Table\ref{tab:drbp_test474_variant}, DRBP recognition is also slightly improved when the gates are incorporated .
	\begin{table}[H]
		\centering
		\begin{tabular}{lcc}
			\toprule
			\textbf{Method} & \textbf{Number of proteins} & \textbf{Number of }\\
			& \textbf{predicted as DRBP} & \textbf{true DRBP }\\
			\midrule
			Variant 4   & 11 & 6\\
			Variant 5  & 9 & 6   \\
			Variant 6 (LAMP-PRo)  & 7 & 6\\
			\bottomrule
		\end{tabular}
		\caption{Performance comparison based on DRBP recognition while including gates and without them in TEST474 dataset}
		\label{tab:drbp_test474_variant}
	\end{table}
	
	Overall, Variant 6 which is the proposed model LAMP-PRo, shows the most competent performance, thereby proving that all the components work together to give better prediction results. 

	\subsection{Comparison with existing methods}
	\subsubsection{Discriminating DBPs and RBPs}
\begin{figure}[H]
		\centerline{
			\includegraphics[height=2.2in,width=3.5in]{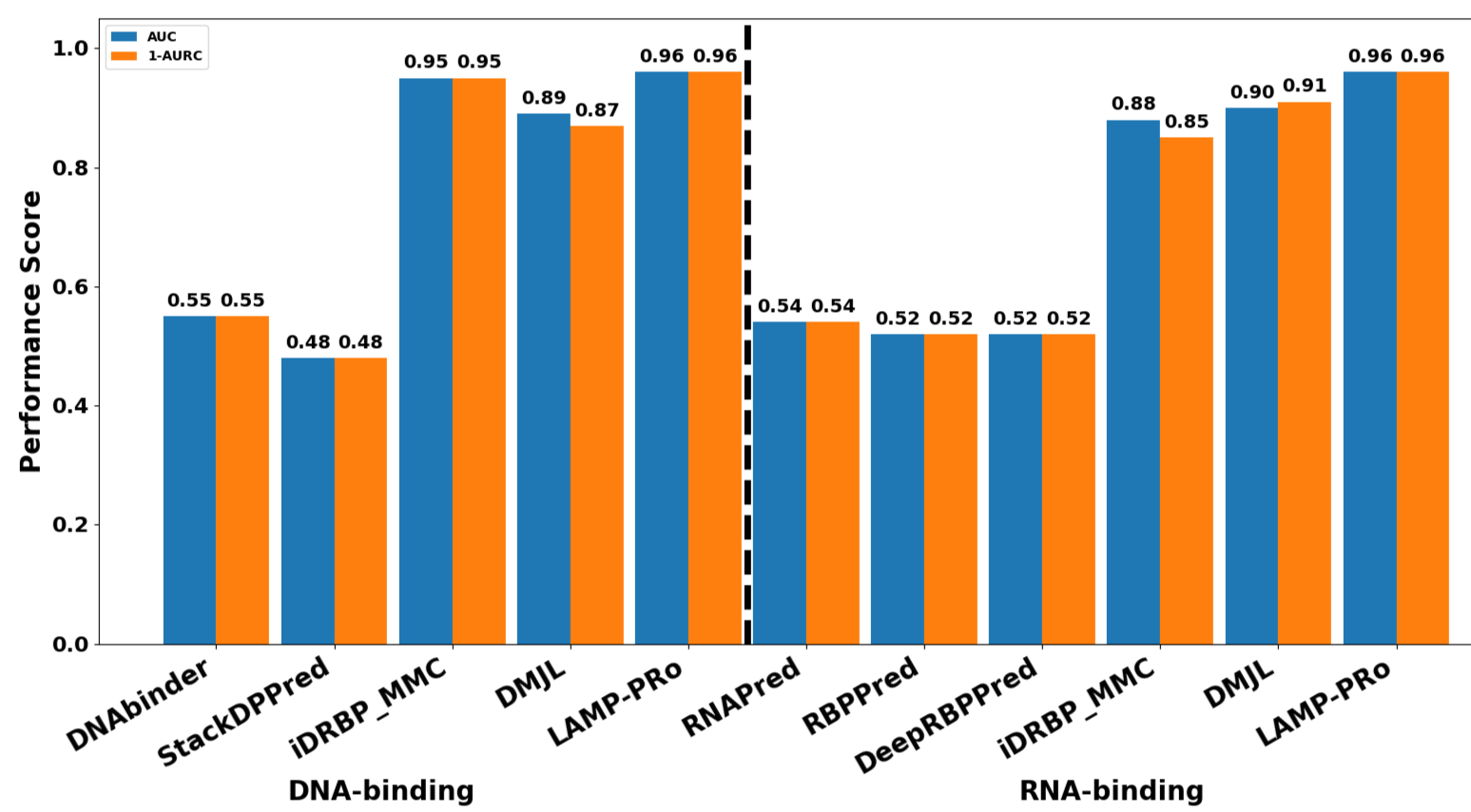}
		}
		\caption{Comparison of different methods on EZL dataset}
		\label{EZL_comparison}
	\end{figure}

	To understand the discriminating ability of LAMP-PRo for DBPs and RBPs, EZL dataset is considered as it contains only DBP and RBP data. Fig.~\ref{EZL_comparison} depicts the comparison of the different methods on EZL dataset. 
	
	As is evident from the figure, DNAbinder and StackDPPred can identify only DBPs while RNAPred, RBPPRed and DeepRBPPred can identify RBPs only leading to very low 1-AURC values. This proves that these methods cannot deal with cross-prediction problems efficiently. On the other hand, though methods like iDRBP\_MMC and DMJL models can deal with cross-prediction problem (1-AURC values for identifying DNA-binding proteins are 0.95 and 0.87 respectively while for identifying RNA-binding proteins such values are 0.85 and 0.91), with an 1-AURC value of 0.96 for identifying both DNA-binding and RNA-binding proteins, the proposed method LAMP-Pro shows better performance than the existing methods. This confirms the fact that LAMP-PRo can identify both DBPs and RBPs well on the EZL dataset, thus able to handle cross-prediction problem much better than the existing methods.
	
	\subsubsection{Binding Protein Recognition}
	Some studies have shown that how proteins binding to both DNA and RNA (DBRPs) can control gene expression better as well as regulate protein activity in cells. We have compared LAMP-PRo with existing methods to verify if our method can identify DRBPs better than the rest; Tables~\ref{tab:drbp_test474_init} and \ref{tab:drbp_test474} show these comparisons. As can be seen in Table~\ref{tab:drbp_test474_init}, compared to  iDRBP\_MMC~\cite{ZHANG2020} and DMJL model \cite{Du2023}, LAMP-PRo can predict more number of true DRBPs. 
	\begin{table}[H]
		\centering
		\begin{tabular}{lcc}
			\toprule
			\textbf{Method} & \textbf{Number of proteins} & \textbf{Number of }\\
			& \textbf{predicted as DRBP} & \textbf{true DRBP }\\
			\midrule
			iDRBP\_MMC \cite{ZHANG2020}  & 20 & 2\\
			DMJL model \cite{Du2023} & 8 & 2   \\
			LAMP-PRo   & 7 & 6\\
			\bottomrule
		\end{tabular}
		\caption{Performance comparison of LAMP-PRo with iDRBP\_MMC and DMJL Model in terms of DRBP recognition in TEST474 dataset}
		\label{tab:drbp_test474_init}
	\end{table}
	This is also further evident from Table~\ref{tab:drbp_test474}.  As compared to SERCNN~\cite{Prabhu2025}, the recall, precision and F1-score of our method are much better. The values of such parameters are 0.75, 0.60 and 0.667 as compared to SERCNN's 0.5, 0.16 and 0.242, showing an improvement of around 50\% for recall and more than 100\% for both precision and F1-score. This result can be further elucidated by the number of DRBPs correctly identified by LAMP-PRo. There are 8 DRBPs in the data out of which our model could identify 6 correctly. On the other hand, SERCNN could presumably only identify 4. The results depict that our method performs better than all the existing methods in identifying DRBPs.
	\begin{table}[H]
		\centering
		\begin{tabular}{lccc}
			\toprule
			\textbf{Method} & \textbf{Recall} & \textbf{Precision} & \textbf{F1-Score} \\
			\midrule
			iDRBP\_MMC \cite{ZHANG2020}  & 0.25  & 0.143 & 0.182 \\
			DMJL model \cite{Du2023} & 0.25  & 0.25  & NA    \\
			iDRBP-EL \cite{Wang2023}    & 0.375 & 0.158 & 0.222 \\
			SERCNN  \cite{Prabhu2025}           & 0.500  & 0.160  & 0.242 \\
			LAMP-PRo   & 0.75 & 0.85 & 0.80\\
			\bottomrule
		\end{tabular}
		\caption{Performance comparison of LAMP-PRo with existing methods for predicting DRBP in TEST474 dataset}
		\label{tab:drbp_test474}
	\end{table}
	To further evaluate the ability of LAMP-PRo for DRBP recognition, our model is tested on DRBP206 dataset which consists of only DRBP and Non-NABP dataset. The results of this evaluation are provided in Table~\ref{tab:drbp_DRBP206}. As can be seen from the table, with an AUC, accuracy and MCC of 0.96, 0.88 and 0.79 respectively, LAMP-PRo outperforms all the existing methods. These results further confirm the superiority of LAMP-PRo in recognizing DRBP which is much better than the existing methods.  
	\begin{table}[H]
		\centering
		
		\begin{tabular}{lccc}
			\toprule
			\textbf{Method} & \textbf{AUC} & \textbf{Accuracy} & \textbf{MCC} \\
			\midrule
			iDRBP\_MMC \cite{ZHANG2020}  & 0.71  & 0.64 & 0.32 \\
			DMJL model \cite{Du2023}    & 0.69 & 0.66 & 0.34 \\
			SERCNN  \cite{Prabhu2025}           & 0.74  & 0.69 & 0.38 \\
			LAMP-PRo   & 0.96 & 0.88 & 0.79\\
			\bottomrule
		\end{tabular}
		\caption{Performance comparison of our model with existing methods for predicting DNA and RNA in DRBP206 dataset}
		\label{tab:drbp_DRBP206}
	\end{table}
	
	\subsubsection{Independent Tests}
	The performance of the proposed LAMP-PRo model is compared with several existing state-of-the-art predictors which include models like DNAbinder \cite{Kumar2007}, StackDPPred \cite{Mishra2018}, RNA Pred \cite{Kumar2011}, RBPPred \cite{Zhang2016},  Deep-RBPPred \cite{Zheng2018}, iDRBP\_MMC \cite{ZHANG2020}, iDRBP-EL \cite{Wang2023}, Dmjl model \cite{Du2023} and SERCNN~\cite{Prabhu2025}. For an overall comparison, TEST474 and PDB255 are chosen as the independent test datasets. The results are depicted in Table~\ref{tab:binding_results}. For DNA Binding, LAMP-PRo showcases the best performance with an AUC of 0.98 and 1-AURC of 0.97 (showing improvements of 3\% and 6\% respectively than the best state-of-the art) for TEST474 and an AUC of 0.80 for PDB255. 
	LAMP-PRo also achieves an 1-AURC value of 0.95 for RNA Binding prediction on TEST474 dataset while for PDB255, such prediction fetches an AUC and 1-AURC of 0.83.
	While LAMP-PRo achieves the highest AUC and 1-AURC on most benchmarks, its RNA binding performance in terms of AUC on TEST474 and DNA binding performance on PDB255 are slightly lower than some baselines. This can be attributed to the model’s focus on multi-label and DRBP prediction, where it effectively captures label-specific features and dependencies between DBP and RBP, resulting in improved recognition of dual-binding proteins without compromising overall multi-label performance.
	
	\begin{table*}[ht]
		\centering
		\caption{Performance comparison of LAMP-PRo with existing methods on TEST474 and PDB255 datasets for DNA and RNA binding prediction.}
		\begin{tabular}{l|cc|cc|cc|cc}
			\hline
			\multirow{2}{*}{\textbf{Method}} & \multicolumn{4}{c|}{\textbf{TEST474}} & \multicolumn{4}{c}{\textbf{PDB255}} \\\cline{2-9}
			& \multicolumn{2}{c|}{DNA Binding} & \multicolumn{2}{c|}{RNA Binding} & \multicolumn{2}{c|}{DNA Binding} & \multicolumn{2}{c}{RNA Binding} \\\cline{2-9}
			& AUC & 1-AURC & AUC & 1-AURC & AUC & 1-AURC & AUC & 1-AURC \\
			\hline
			DNAbinder \cite{Kumar2007}      & 0.86 & 0.68 & NA   & NA   & 0.54 & 0.44 & NA   & NA   \\
			StackDPPred \cite{Mishra2018}     & 0.76 & 0.54 & NA   & NA   & 0.65 & 0.55 & NA   & NA   \\
			RNA Pred \cite{Kumar2011}       & NA   & NA   & 0.66 & 0.50 & NA   & NA   & 0.75 & 0.75 \\
			RBPPred \cite{Zhang2016}        & NA   & NA   & 0.73 & 0.54 & NA   & NA   & 0.70 & 0.71 \\
			Deep-RBPPred \cite{Zheng2018}   & NA   & NA   & 0.58 & 0.41 & NA   & NA   & 0.72 & 0.75 \\
			iDRBP\_MMC \cite{ZHANG2020}     & 0.87 & 0.82 & 0.80 & 0.76 & 0.70 & 0.69 & 0.74 & 0.78 \\
			iDRBP-EL \cite{Wang2023}        & 0.95 & 0.89 & 0.91 & 0.93 & 0.83 & 0.84 & 0.78 & 0.78 \\
			DMJL model \cite{Du2023}     & 0.94 & 0.87 & 0.83 & 0.85 & 0.77 & 0.82 & 0.72 & 0.71 \\
			SERCNN~\cite{Prabhu2025}                     & 0.95 & 0.91 & 0.92 & 0.90 & 0.80 & 0.85 & 0.80 & 0.81 \\
			LAMP-PRo & 0.98 & 0.97 & 0.90 & 0.95 & 0.80 & 0.78 &0.81 & 0.83\\
			\hline
		\end{tabular}
		\label{tab:binding_results}
	\end{table*}

	\section{Visual Analysis}
\begin{figure*}
		
		\centerline{
			\includegraphics[height=0.8in,width=4.0in]{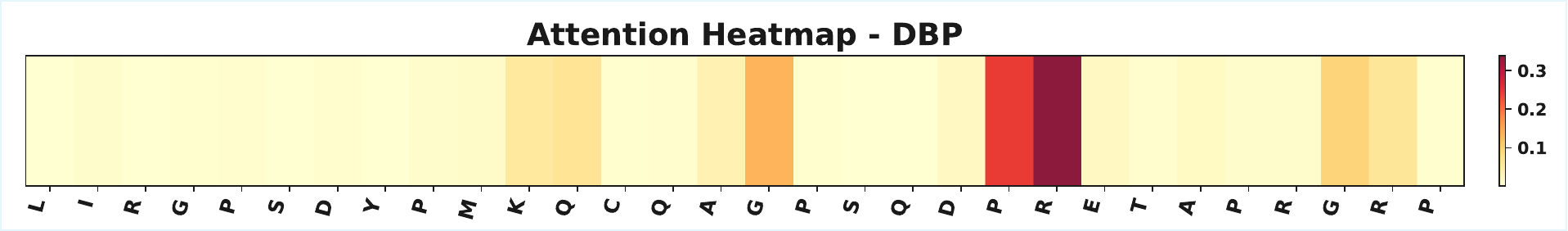}
			\includegraphics[height=0.8in,width=4.0in]{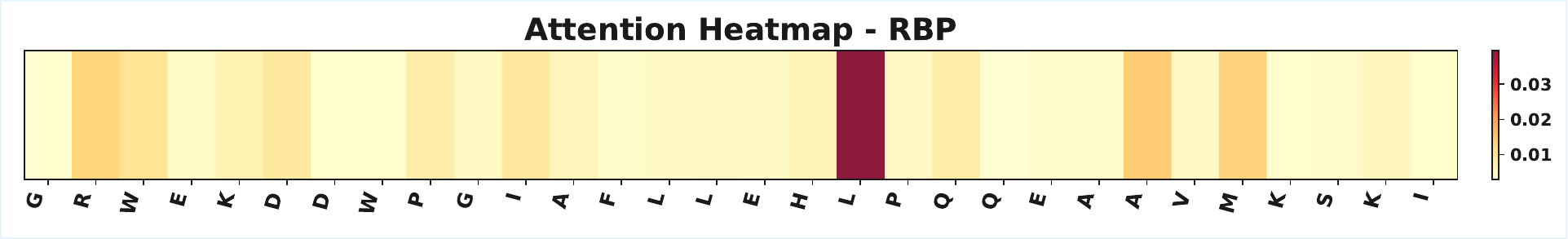}
		}\centerline{(a)\hspace{90mm}(b)}
		\centerline{
			\includegraphics[height=0.8in,width=4.0in]{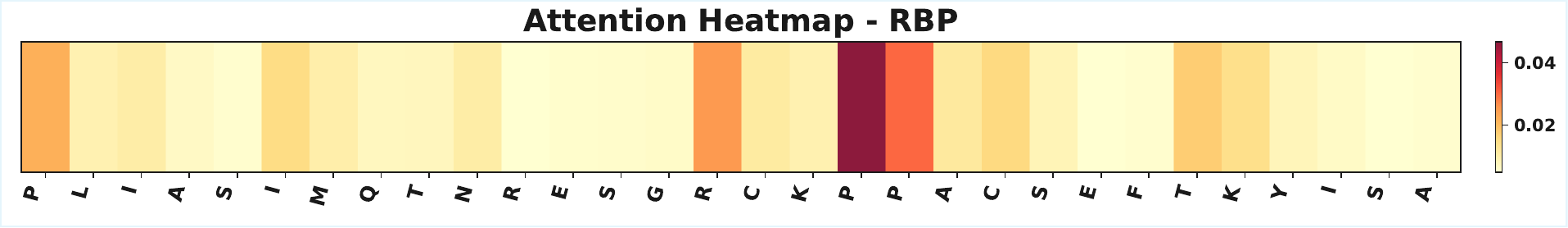}
			\includegraphics[height=0.8in,width=4.0in]{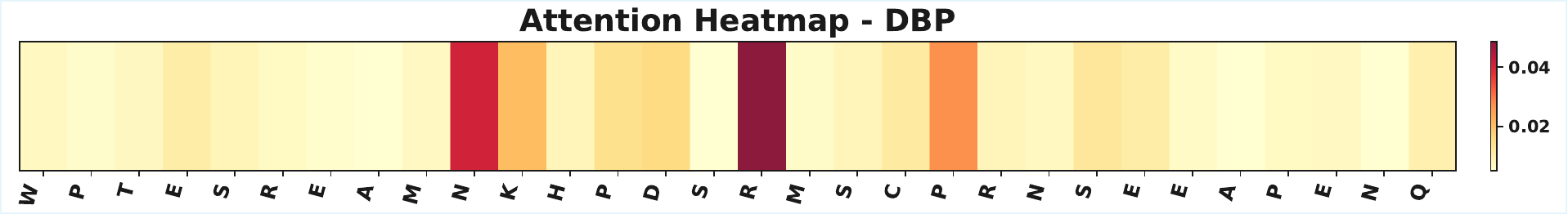}
		}
		\centerline{(c)\hspace{90mm}(d)}
		\caption{Visual analysis of correctly predicted sequences where (a)  DBP  and (b) RBP attention weights for a DBP sequence and (c) RBP and (d) DBP attention weights for a RBP sequence}
		\label{fig:vis}
	\end{figure*}

	To better understand LAMP-PRo's decision process, we predicted the attention weights of the top 30 tokens (amino acids) for the final prediction of two correctly predicted protein sequences; DBP and RBP. In this regard, Fig.~\ref{fig:vis} highlights such residues in sequences.  Fig.~\ref{fig:vis}(a) and (b) are the attention heatmaps for sequences where the figures respectively showcase the attention weights for a sequence correctly predicted as DBP and the corresponding RBP values. In the DBP predicted sequence (Fig.~\ref{fig:vis}(a)), the peak is much stronger ($\sim$0.3) as compared to that of RBP ($\sim$0.03) in Fig.~\ref{fig:vis}(b). This suggests that the model pays more attention to certain residues for DBP than for RBP. Thus, LAMP-PRo assigns a higher confidence score to DBP, thereby leading to a correct prediction.  The attention weights highlight residue like K (Lysine) and R (Arginine) which are frequently involved in DNA-binding domains~\cite{Jones2003}. 
	On the other hand, P (Proline) also shows a moderately high attention weight. While proline  alone may not be a definitive marker for DNA-binding activity, its presence can significantly influence DNA interaction capabilities~\cite{Agarwal2022}. 
	

	Fig.~\ref{fig:vis}(c) and (d) showcase the attention weights for a sequence predicted correctly as RBP and the corresponding DBP values. Although P (proline) shows high attention weights in Fig~\ref{fig:vis}(a) as well
	and it is essential in both RNA and DNA contexts, its roles in RNA are often more dynamic and functionally diverse, encompassing structural, catalytic, and regulatory functions. For instance, the RNA-binding protein YebC enhances translation of proline-rich amino acid stretches in bacteria, highlighting the importance of proline in RNA-binding contexts~\cite{Ignatov2025}. The importance of proline is also evident in Fig.~\ref{fig:vis}(c) where multiple proline residues are present with high attention weights.
	
	It is worth mentioning here that the attention weights are normalized within each sequence but aggregated across heads, and therefore their absolute values may not necessarily sum to 1.
	\subsection{Biological Relevance}
	In order to provide biological insight of LAMP-PRo, all the test datasets are combined leading to a total of 2443 DBPs, 1915 RBPs and 111 DRBPs after removing all the duplicate proteins. Out of such proteins, LAMP-PRo can successfully identify  1957 DBPs ($\sim$80\%), 1633  RBPs ($\sim$85\%) and 85 DRBPs ($\sim$76\%) as opposed to DMJL model which can only identify 1921 DBPs, 1327 RBPs and 14 DRBPs. Many of the proteins identified by LAMP-PRo are involved in various diseases like cancer, intellectual disability and neurodegenerative  disorder. In this regard, some proteins related to cancer are FUS (ID: P35637), RAR-alpha (ID: P10276) and NONO (ID: Q15233). FUS is related to acute myeloid leukemia (AML)~\cite{Ichikawa1994} while chromosomal aberrations involving RARA and NONO are commonly found in acute promyelocytic leukemia~\cite{Fujita2003} and papillary renal cell carcinoma~\cite{Clark1997} respectively. FUS (ID: P35637) and TABP (ID: Q13148) are related to a specific neurodegenerative disorder called amyotrophic lateral sclerosis which affects motor neurons in the cortex, brain stem and spinal cord~\cite{Hou2016}. Others like EP300 (ID: Q09472), DDX3X (ID: O00571), AGO2 (ID: Q9UKV8), DHX9 (ID: Q08211), AGO1 (ID: Q9UL18), HNRPK (ID: P61978), RBMX (ID: P38159), H14 (ID: P10412), SON (ID: P18583) and TYY1 (ID: P25490) are associated with different forms of intellectual disabilities. For example, AGO1~\cite{SAKAGUCHI2019} and DHX9~\cite{YAMADA2023} are involved in autism spectrum disorder. From the above discussion, we can conclude that LAMP-PRo can successfully identify several proteins related to different diseases.

	\subsection{Performance of LAMP-PRo for identifying new DBPs and RBPs}
	To show the effectiveness of LAMP-PRo for identifying new nucleic acid-binding proteins, we have used 36 DBPs and 4 RBPs available at  http://bliulab.net/iDRBP-EL. These are reviewed proteins according to GO terms collected from the Swiss-Prot database. LAMP-PRo correctly identifies 34 of the 36 DBPs and all the 4 RBPS, thereby proving that it can accurately predict new proteins. One of the identified DBPs ATF7\_CAEEL (ID: Q86MD3), acts as a master switch that turns on genes during immune defense in Caenorhabditis elegans as well as protects the organism from oxidative stress~\cite{Fletcher2019, Hall2017}. Thus this protein can be considered to be critical for protecting the organism's cells and maintaining health. FB11A\_DANRE (ID: A0A2R8QFQ6) is an identified RBP which plays a key role in brain, eye and jaw development in Zebrafish \cite{HOLT2019}.
	
	\section{Conclusion}
	Cross-prediction is a common problem while identifying DNA-binding (DBPs) and RNA-binding (RBPs) proteins. Furthermore, identifying proteins binding to both DNA and RNA (DBRP) pose another challenge. To mitigate such problems, in this work we have proposed LAMP-PRo which utilizes the benefits of ESM-2 embeddings, CNN, multi-head self-attention (MHSA), label-aware attention (LAA) and cross-label attention (CLA). Rich  and informative protein representations are contributed by ESM-2 while CNN provides a high-order local pattern extraction. On the other hand, MHSA captures the global dependencies between the amino acids and label-aware attention learns the specific attention patterns per label, thereby helping to reduce cross-prediction problem. Finally, to  improve the prediction of DBRP, cross-label attention is incorporated in the model. The benefit of all these components are shown through ablation studies. Furthermore, we have also included a gated residual connection between CNN and MHSA as well as between LAA and
	CLA to adaptively regulate local-global
	feature fusion and stabilize training. Moreover, we have also introduced invalid label penalty to discourage biologically inconsistent label combinations. Incorporating all these features contribute to the competent performance of LAMP-Pro which is showcased by comparing with existing models where the model is shown to perform consistently well. Although, LAMP-PRo shows competent performance as compared to the existing models, our results demonstrate that some components like gated residual connection are not universally beneficial across all datasets. Instead, their impact depends on dataset label distribution as well as task complexity. These findings highlight the importance of tailoring model complexity to dataset characteristics, and we release our framework openly to enable the community to explore these trade-offs further. 
	
	Although, LAMP-PRo shows consistently good performance as compared to the existing models, we would like to further improve its predictive ability by incorporating epigenomics, transcriptomics and structural genomics data.
	\section{Declaration of competing interest}
	The authors do not have any competing interest to declare.
	\bibliographystyle{elsarticle-num}
	\bibliography{Ref}

\begin{thebibliography}{10}
\expandafter\ifx\csname url\endcsname\relax
  \def\url#1{\texttt{#1}}\fi
\expandafter\ifx\csname urlprefix\endcsname\relax\def\urlprefix{URL }\fi
\expandafter\ifx\csname href\endcsname\relax
  \def\href#1#2{#2} \def\path#1{#1}\fi

\bibitem{Du2023}
X.~Du, J.~Hu, Deep multi-label joint learning for {RNA} and {DNA}-binding
  proteins prediction, IEEE/ACM Transactions on Computational Biology and
  Bioinformatics 20~(1) (2023) 307--320.
\newblock \href {https://doi.org/10.1109/TCBB.2022.3150280}
  {\path{doi:10.1109/TCBB.2022.3150280}}.

\bibitem{Prabhu2025}
T.~P. Prabhu, R.~Bethi, S.~Mahapatra, et~al., Attention integrated residue cnn
  for classification of {DNA}-binding and {RNA}-binding proteins, IEEE Access
  13 (2025) 96226--96235.
\newblock \href {https://doi.org/10.1109/ACCESS.2025.3575700}
  {\path{doi:10.1109/ACCESS.2025.3575700}}.

\bibitem{Cui2022}
M.~Cui, C.~Cheng, L.~Zhang, High-throughput proteomics: a methodological
  mini-review, Laboratory Investigation 102 (2022) 1170–1181.
\newblock \href {https://doi.org/10.1038/s41374-022-00830-7}
  {\path{doi:10.1038/s41374-022-00830-7}}.

\bibitem{Douglas2007}
S.~M. Douglas, J.~J. Chou, W.~M. Shih, Dna-nanotube-induced alignment of
  membrane proteins for nmr structure determination, Proceedings of the
  National Academy of Sciences 104~(16) (2007) 6644--6648.
\newblock \href {https://doi.org/10.1073/pnas.0700930104}
  {\path{doi:10.1073/pnas.0700930104}}.

\bibitem{Jaiswal2015}
R.~Jaiswal, S.~K. Singh, D.~Bastia, et~al., {Crystallization and preliminary
  X-ray characterization of the eukaryotic replication terminator Reb1{--}Ter
  DNA complex}, Acta Crystallographica Section F 71~(4) (2015) 414--418.
\newblock \href {https://doi.org/10.1107/S2053230X15004112}
  {\path{doi:10.1107/S2053230X15004112}}.

\bibitem{Kumar2007}
M.~Kumar, M.~M. Gromiha, G.~P. Raghava, Identification of dna-binding proteins
  using support vector machines and evolutionary profiles, BMC Bioinformatics
  8~(463) (2007).
\newblock \href {https://doi.org/10.1186/1471-2105-8-463}
  {\path{doi:10.1186/1471-2105-8-463}}.

\bibitem{Mahmud2024}
S.~M.~H. Mahmud, M.~F.~H. K.~O. M.~Goh, et~al., {Deep-WET}: a deep
  learning-based approach for predicting {DNA}-binding proteins using word
  embedding techniques with weighted features, Scientific Reports 14~(2961)
  (2024).
\newblock \href {https://doi.org/10.1038/s41598-024-52653-9}
  {\path{doi:10.1038/s41598-024-52653-9}}.

\bibitem{Ahmed2024}
S.~H. Ahmed, D.~B. Bose, R.~Khandoker, et~al., Stackdpp: a stacking ensemble
  based dna-binding protein prediction model, BMC Bioinformatics 25~(111)
  (2024).
\newblock \href {https://doi.org/10.1186/s12859-024-05714-9}
  {\path{doi:10.1186/s12859-024-05714-9}}.

\bibitem{ZENG2024}
W.~Zeng, X.~Yu, J.~Shang, et~al., {LBi-DBP}, an accurate dna-binding protein
  prediction method based lightweight interpretable bilstm network, Expert
  Systems with Applications 249 (2024) 123525.
\newblock \href {https://doi.org/10.1016/j.eswa.2024.123525}
  {\path{doi:10.1016/j.eswa.2024.123525}}.

\bibitem{Kumar2011}
M.~Kumar, M.~Gromiha, G.~Raghava, {SVM} based prediction of rna-binding
  proteins using binding residues and evolutionary information, Journal of
  molecular recognition 24 (2011) 303--13.
\newblock \href {https://doi.org/10.1002/jmr.1061}
  {\path{doi:10.1002/jmr.1061}}.

\bibitem{Ma2015}
X.~Ma, J.~Guo, K.~Xiao, et~al., {PRBP}: Prediction of rna-binding proteins
  using a random forest algorithm combined with an rna-binding residue
  predictor, IEEE/ACM Transactions on Computational Biology and Bioinformatics
  12~(6) (2015) 1385--1393.
\newblock \href {https://doi.org/10.1109/TCBB.2015.2418773}
  {\path{doi:10.1109/TCBB.2015.2418773}}.

\bibitem{Zhang2016}
X.~Zhang, S.~Liu, {RBPP}red: predicting {RNA}-binding proteins from sequence
  using {SVM}, Bioinformatics 33~(6) (2016) 854--862.
\newblock \href {https://doi.org/10.1093/bioinformatics/btw730}
  {\path{doi:10.1093/bioinformatics/btw730}}.

\bibitem{Zheng2018}
J.~Zheng, X.~Zhang, X.~Zhao, et~al., {Deep-RBPPred}: Predicting rna binding
  proteins in the proteome scale based on deep learning, Scientific Reports
  8~(15264) (10 2018).
\newblock \href {https://doi.org/10.1038/s41598-018-33654-x}
  {\path{doi:10.1038/s41598-018-33654-x}}.

\bibitem{Yan2024}
Y.~Yan, W.~Li, S.~Wang, et~al., {Seq-RBPPred}: Predicting rna-binding proteins
  from sequence, ACS Omega 9~(11) (2024) 12734--12742.
\newblock \href {https://doi.org/10.1021/acsomega.3c08381}
  {\path{doi:10.1021/acsomega.3c08381}}.

\bibitem{Zhang2021}
J.~Zhang, , Q.~Chen, B.~Liu, {DeepDRBP-2L}: A new genome annotation predictor
  for identifying dna-binding proteins and rna-binding proteins using
  convolutional neural network and long short-term memory, IEEE/ACM
  Transactions on Computational Biology and Bioinformatics 18~(4) (2021)
  1451--1463.
\newblock \href {https://doi.org/10.1109/TCBB.2019.2952338}
  {\path{doi:10.1109/TCBB.2019.2952338}}.

\bibitem{Wu2024}
S.~Wu, J.~t.~Guo, Improved prediction of dna and rna binding proteins with deep
  learning models, Briefings in Bioinformatics 25~(4) (2024) bbae285.
\newblock \href {https://doi.org/10.1093/bib/bbae285}
  {\path{doi:10.1093/bib/bbae285}}.

\bibitem{ZHANG2020}
J.~Zhang, Q.~Chen, B.~Liu, {iDRBP\_MMC}: Identifying {DNA}-binding proteins and
  {RNA}-binding proteins based on multi-label learning model and motif-based
  convolutional neural network, Journal of Molecular Biology 432~(22) (2020)
  5860--5875.
\newblock \href {https://doi.org/10.1016/j.jmb.2020.09.008}
  {\path{doi:10.1016/j.jmb.2020.09.008}}.

\bibitem{Wang2023}
N.~Wang, J.~Zhang, B.~Liu, {iDRBP-EL}: Identifying {DNA- and RNA-} binding
  proteins based on hierarchical ensemble learning, IEEE/ACM Transactions on
  Computational Biology and Bioinformatics 20~(1) (2023) 432--441.
\newblock \href {https://doi.org/10.1109/TCBB.2021.3136905}
  {\path{doi:10.1109/TCBB.2021.3136905}}.

\bibitem{Mishra2018}
A.~Mishra, P.~Pokhrel, M.~T. Hoque, Stackdppred: a stacking based prediction of
  dna-binding protein from sequence, Bioinformatics 35~(3) (2018) 433--441.
\newblock \href {https://doi.org/10.1093/bioinformatics/bty653}
  {\path{doi:10.1093/bioinformatics/bty653}}.

\bibitem{Jones2003}
S.~Jones, H.~P. Shanahan, H.~M. Berman, et~al., Using electrostatic potentials
  to predict dna‐binding sites on dna‐binding proteins, Nucleic Acids
  Research 31~(24) (2003) 7189--7198.
\newblock \href {https://doi.org/10.1093/nar/gkg922}
  {\path{doi:10.1093/nar/gkg922}}.

\bibitem{Agarwal2022}
N.~Agarwal, N.~Nagar, R.~Raj, et~al., Conserved apical proline regulating the
  structure and dna binding properties of helicobacter pylori histone-like dna
  binding protein (hup), ACS Omega 7~(17) (2022) 15231--15246.
\newblock \href {https://doi.org/10.1021/acsomega.2c01754}
  {\path{doi:10.1021/acsomega.2c01754}}.

\bibitem{Ignatov2025}
D.~Ignatov, V.~Shanmuganathan, R.~A.-Begrich, et~al., Rna-binding protein yebc
  enhances translation of proline-rich amino acid stretches in bacteria, Nature
  Communications 16~(6262) (2025).
\newblock \href {https://doi.org/10.1038/s41467-025-60687-4}
  {\path{doi:10.1038/s41467-025-60687-4}}.

\bibitem{Ichikawa1994}
H.~Ichikawa, K.~Shimizu, Y.~Hayashi, et~al., An rna-binding protein gene,
  tls/fus, is fused to erg in human myeloid leukemia with t(16;21) chromosomal
  translocation, Cancer research 54~(11) (1994) 2865—2868.

\bibitem{Fujita2003}
K.~Fujita, R.~Oba, H.~Harada, et~al., Cytogenetics, fish and rt-pcr analysis of
  acute promyelocytic leukemia: Structure of the fusion point in a case lacking
  classic t(15;17) translocation, Leukemia \& Lymphoma 44~(1) (2003) 111--115.
\newblock \href {https://doi.org/10.1080/1042819021000040305}
  {\path{doi:10.1080/1042819021000040305}}.

\bibitem{Clark1997}
J.~Clark, Y.~Lu, S.~Sidhar, et~al., Fusion of splicing factor genes psf and
  nono (p54nrb) to the tfe3 gene in papillary renal cell carcinoma, Oncogene
  15~(18) (1997) 2233—2239.
\newblock \href {https://doi.org/10.1038/sj.onc.1201394}
  {\path{doi:10.1038/sj.onc.1201394}}.

\bibitem{Hou2016}
L.~Hou, B.~Jiao, T.~Xiao, et~al., Screening of {SOD1}, {FUS} and {TARDBP} genes
  in patients with amyotrophic lateral sclerosis in central-southern china,
  Scientific Reports 6~(32478) (2016).
\newblock \href {https://doi.org/10.1038/srep324784}
  {\path{doi:10.1038/srep324784}}.

\bibitem{SAKAGUCHI2019}
A.~Sakaguchi, Y.~Yamashita, T.~Ishii, et~al., Further evidence of a causal
  association between {AGO1}, a critical regulator of micro{RNA} formation, and
  intellectual disability/autism spectrum disorder, European Journal of Medical
  Genetics 62~(6) (2019) 103537.
\newblock \href {https://doi.org/10.1016/j.ejmg.2018.09.004}
  {\path{doi:10.1016/j.ejmg.2018.09.004}}.

\bibitem{YAMADA2023}
M.~Yamada, Y.~Nitta, T.~Uehara, et~al., Heterozygous loss-of-function {DHX9}
  variants are associated with neurodevelopmental disorders: Human genetic and
  experimental evidences, European Journal of Medical Genetics 66~(8) (2023)
  104804.
\newblock \href {https://doi.org/10.1016/j.ejmg.2023.104804}
  {\path{doi:10.1016/j.ejmg.2023.104804}}.

\bibitem{Fletcher2019}
M.~Fletcher, E.~J. Tillman, V.~L. Butty, et~al., Global transcriptional
  regulation of innate immunity by {ATF-7} in {C. elegans}, PLOS Genetics
  15~(2) (2019) 1--14.
\newblock \href {https://doi.org/10.1371/journal.pgen.1007830}
  {\path{doi:10.1371/journal.pgen.1007830}}.

\bibitem{Hall2017}
J.~A. Hall, M.~K. McElwee, J.~H. Freedman, Identification of {ATF-7} and the
  insulin signaling pathway in the regulation of metallothionein in {C.
  elegans} suggests roles in aging and reactive oxygen species, PLOS ONE 12~(6)
  (2017) 1--22.
\newblock \href {https://doi.org/10.1371/journal.pone.0177432}
  {\path{doi:10.1371/journal.pone.0177432}}.

\bibitem{HOLT2019}
R.~J. Holt, R.~M. Young, B.~Crespo, et~al., De novo missense variants in
  {FBXW11} cause diverse developmental phenotypes including brain, eye, and
  digit anomalies, The American Journal of Human Genetics 105~(3) (2019)
  640--657.
\newblock \href {https://doi.org/10.1016/j.ajhg.2019.07.005}
  {\path{doi:10.1016/j.ajhg.2019.07.005}}.

\end{thebibliography}
\end{document}